\documentclass[graybox]{svmult}
\AtBeginDocument{%
  \providecommand\BibTeX{{%
    \normalfont B\kern-0.5em{\scshape i\kern-0.25em b}\kern-0.8em\TeX}}}

\usepackage{mdframed}
\usepackage{graphicx}
\usepackage{hyperref}
\usepackage{multicol} 

\begin{document}

\title{Teaching Action Research}

\author{Miroslaw Staron}
\institute{Miroslaw Staron \at Computer Science and Engineering, Chalmers $|$ University of Gothenburg, \email{miroslaw.staron@gu.se}}

\maketitle

\abstract{
Action research entered into software engineering as one of the responses to the software engineering research crisis at the end of the last millennium. As one of the challenges in the crisis was the lack of empirical results and the transfer of research results into practices, the action research could address these challenges. It is a methodology where collaboration and host organizations are the focus of knowledge discovery, development, and documentation. Although the methodology is often well received in industrial contexts, it is inherently difficult to learn as it requires experience and varies from organization to organization. This chapter describes the pillars of action research as a methodology and how to teach them. The chapter includes examples of teaching action research at the bachelor, master and PhD levels. In addition to theory, the chapter contains examples from practice. }




\newmdenv[
roundcorner=50pt,
subtitlebelowline=true,subtitleaboveline=true,
subtitlebackgroundcolor=gray!50!white,
backgroundcolor=gray!20!white,
frametitle={Example},frametitlerule=true,
frametitlebackgroundcolor=gray!70!white,
]{subtitleenv} 

\newmdenv[
roundcorner=100pt,
subtitlebelowline=true,subtitleaboveline=true,
subtitlebackgroundcolor=gray!50!white,
backgroundcolor=gray!10!white,
frametitle={Important},frametitlerule=true,
frametitlebackgroundcolor=gray!40!white,
]{importantenv}

\newmdenv[
roundcorner=100pt,
subtitlebelowline=true,subtitleaboveline=true,
subtitlebackgroundcolor=gray!20!white,
backgroundcolor=gray!10!white,
frametitle={Teaching guideline},frametitlerule=true,
frametitlebackgroundcolor=gray!40!white,
]{teachingenv}

\section{Introduction}
\noindent 
The epistemology of action research is rooted in knowledge being a product of collaboration and context. It stems from doing rather than observing or testing. The knowledge developed in action research studies is about improving existing practices by combining these practices with new interventions. When engaging in action research projects, we intend to introduce changes to operations of the host organizations; we want to observe what these interventions (actions) have for consequences and we want to learn from them. 

With this epistemological stance, this research methodology is quite unique as it combines constructive research with observational participatory studies. Action research is defined by Reason and Bradbury as \emph{a participatory,
democratic process concerned with developing practical knowing in the pursuit of worthwhile human purposes, grounded in a participatory worldview which we believe is emerging at this historical moment. It seeks to bring together action and reflection, theory and practice, in participation with others, in the pursuit of practical solutions to issues of pressing concern to people, and more generally the flourishing of individual persons and their communities} \cite{reason2001handbook}. It emphasizes the source of knowledge as originating from contemporary problems and the collaboration between participants of the research study. 

This is a view that has evolved over the last three decades from the definition of Participatory Action Research  \cite{argyris1989participatory}: \emph{a form of action research that involves practitioners as both subjects and coresearchers.} It has been adopted in information systems by  \cite{baskerville1999action} in the end of 1990s. It is also the same definition that we use in software engineering \cite{staron2020action}. 

Action research is a methodology that complements experiments, case studies, surveys and design science research methodologies in software engineering. It has its own merits in its flexibility of the design and learning as well as intervening in practitioners' work. Scientists appreciate the ability to be part of host organizations while conducting their studies, getting the ability to understand the details and peculiarities of the host organizations. Practitioners value the applied nature of the research conducted with their academic partners.  

Action research is also flexible, as it allows the action research team to adjust to the situation in the host organization. The flexible nature of action research allows research projects to adjust over time and therefore sustain collaborations over years \cite{sandberg2011agile}. As the host organization evolves, the action research teams can change the pursued research questions and learn new things, taking advantage of new opportunities. As the research field evolves, the host organizations can try novel solution to existing problems, taking advantage of the established trust in the action research team and within the organization. 

However, the advantages of action research are burdened by certain risks. Since the research studies are embedded in organizations, the action research team needs to understand the generalizability of their findings. The results must be transferable to other organizations and contexts. The transparency of the results must be balanced with their generalizability and the security of the company. 

It is because of the flexible nature of the action research that we have inherent problems in understanding it, teaching it to new researchers, and applying it in organizations that operate in highly competitive markets (like AI, telecom, defense and automotive). First of all, it is difficult to theoretically explain how a good collaboration looks like. Concepts like trust, access to infrastructure, management commitment or company politics are inherently difficult to explain in theory. One needs to have a first-hand experience to personalize the theoretical knowledge about it. Second, a good action research project is done in cycles, which require time. Long periods are better than short as they stimulate building trust and commitment, but are also difficult to achieve as reorganizations, promotions, and changing jobs is frequent in our field. Finally, teaching action research is difficult because every organization, collaboration and country are different in their context, prerequisites and legal frameworks. 

In this chapter we learn about action research studies which cross boundaries of academia and industry. Our focus is on collaborations where at least a part of the action team is not employed by the host organization. In practice, this means that in such set-up, the action team is composed of both academic and industrial researchers understand and practitioners, as opposed to an internal action research project where the researchers and practitioners come from within the organization. 

This chapter is structured as follows. First, we explore the pillars of action research, where we focus on the iterative nature of action research and the interaction between organizations and researchers. In Section \ref{sec:phases} we define each phase of action research and explain what the expected outcomes from these phases are. Since the interaction is done as a team, we then move towards the action research teams themselves in Section \ref{sec:teams}. Then, in Section \ref{sec:host} we discuss characteristics of a good host organization and how to interact with it. The teams are responsible for making interventions which are the pivotal element of any action research project. We focus on the interventions in Section \ref{sec:interventions}, where we discuss what is a good intervention and what is not. We also dive into the question of ethics of action research in Section \ref{sec:ethics}. Finally, we conclude the chapter with a set of guidelines on how to teach action research in Section \ref{sec:guidelines}.

Let's dive into the topic and define the pillars of action research -- what makes it so specific? But before that, let us start with the first teaching guideline.

\begin{teachingenv}
    Action research's practical and industrial nature means that we need to teach it in a specific way. I recommend using the case-based approach advocated by Harvard business school.

    To introduce the action research I recommend reading and discussing one of the existing articles where action research has been used, e.g., \cite{dittrich2008cooperative}. The following aspects should be discussed:
    \begin{itemize}
        \item What is the goal of the study?
        \item What is intervention/action in the study?
        \item How is the collaboration between the researchers and the host company organized?
        \item Which data collection methods are used in this article?
    \end{itemize}

    The discussion should focus on the connection between research and practice; it should be about how the knowledge is produced and where state-of-the-art research meets industrial practices. 
\end{teachingenv}

\section{Pillars of Action Research}
\label{sec:pillars}
\begin{figure}[!htb]
    \centering
    \includegraphics[width=0.8\textwidth]{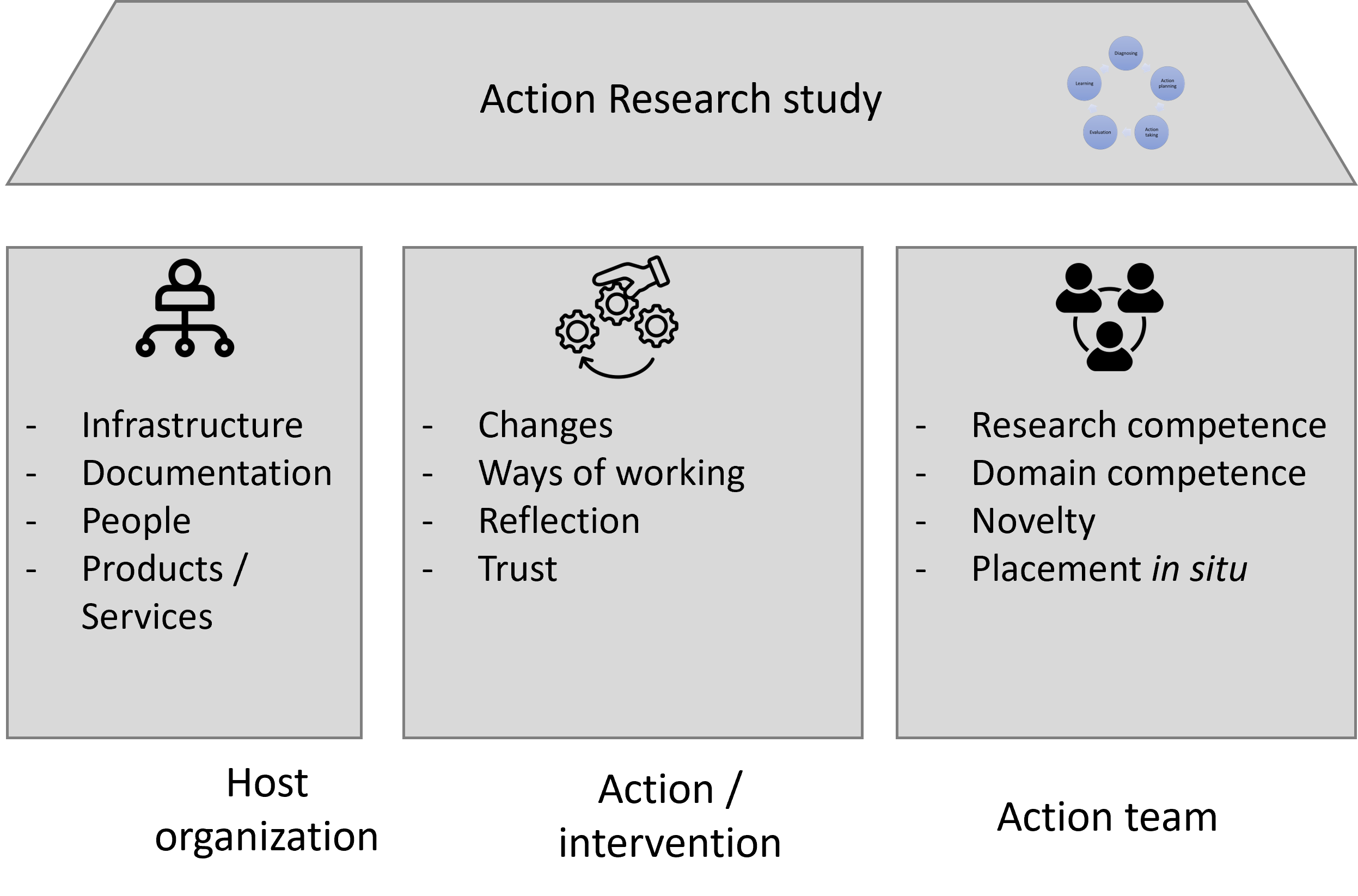}
    \caption{Pillars of action research -- embedding in the host organization, conducting actions/interventions and collaborative, cross-organization action teams.}
    \label{fig:pillars}
\end{figure}

\noindent
Action research is centered around the \textbf{interventions or actions\footnote{In this chapter I use the term action and intervention interchangeably. Although action research has \emph{action} in the name, I often prefer the term \emph{intervention} to emphasize that the action must have an effect on the host organization. I do this to avoid confusion between actions in action research with other types of actions that do not necessarily are interventions.}} -- see Figure \ref{fig:pillars}. The interventions are the part where we do something in order to learn from it. Without the intervention there is no knowledge production, but the interventions are not the only element that is important for the action research project -- we need to prepare for our intervention. 

Without actions/interventions, there is no action research as we do not make any changes or adjustments to how the organizations work. In action research, knowledge is derived from observing the effects of these actions/interventions. 

Actions/interventions must be done \emph{in situ} and therefore the \textbf{host organizations} are equally important. The host organizations must be prepared for entering the action research projects and they must understand the consequences of it. The consequences being that there are inherent risks in changing processes, methods, tools or operations of the host organization. Without them, there is no action research, because we must conduct our actions somewhere.  

Finally, the \textbf{action team} is a special kind of team that includes both researchers and practitioners with distinct roles. They must be part of one team and they must see this team as "their" team in the sense that both the researchers and the practitioners are inside the team. 

The researchers must be an internal part of the team in order to be able to assist in conducting interventions at the host organization. If not possible, they can take peripheral role when taking the intervention (e.g., when it is not possible for the researcher to directly participate in such activities as programming, designing, etc.). They must understand and experience the host organization with its decision structures (formal and informal), processes, products and customers. Only then the researchers can understand which actions can be ethically conducted and which should be ceased. 

The practitioners must also see the action team as part of their organization as they must conduct the interventions. Planning, conducting and evaluating interventions must be included in the work of the practitioners in order to be effective. 

\begin{importantenv}
Action research is not the only research methodology that focuses on collaboration with industry and introduction new tools. Design science research \cite{wieringa2014design} is another one. 

Although these two methodologies share certain characteristics, their epistemological stance on the knowledge creation is very different. 

In action research, the focus is on the three pillars in Figure \ref{fig:pillars}: intervention, action team and the host organization. In design science research, the focus is on the artefact that is being designed, its function and usability. 

When teaching action research, it is important to contrast these two methodologies and explain that the knowledge produced in the action research projects does not have to be in form of new designs or artefacts, but can be about processes, organizations, tools and many more. When teaching action research I provide my students with the following rule of thumb:

\smallskip
\emph{If you can pivot on the solution, you are doing action research; if you can pivot on the organization, then you are using design science research.}
\smallskip

In practice, this means that in action research, we are not bound to a specific tool that needs to be used or evaluated. We can change the focus on an action research cycle and focus on the ways-of-working rather than tools used at the host organization. However, if we must focus on the artifact, then we cannot pivot (change) the focus, but we can change the organization which we collaborate with in order to evaluate our tool, which makes it a design science research cycle. 
\end{importantenv}

\subsection{Knowledge produced during action research}
\noindent
In action research, the knowledge that is produced during the project can be of three different types (mainly):
\begin{enumerate}
    \item Process knowledge -- development of new methods for developing, testing and designing software and ways of working at the host organization \cite{dittrich2008cooperative, dittrich2002doing}. 
    \item Organizational knowledge -- development of new ways of organizing work at the host organization \cite{ferreira2012agile}.
    \item Construction/tooling knowledge -- development of new ways of designing or evaluating software using tools \cite{ochodek2022chapter, calikli2018measure}.
    \item Artefact/models knowledge -- development of new artefacts or models that improve the organization's ways of working.
\end{enumerate}

In action research studies, the introduction of new methods and tools provides two different opportunities. The first one is the ability of the researchers to test new ideas in practice. The second is the ability to have custom-made methods introduced to the industrial practice. 

The latter -- custom-made methods -- allows us to bridge the gap between academic software engineering and the industrial practice \cite{dos2011action}.


\begin{teachingenv}
    When teaching the epistemology of knowledge in software engineering, I find the best tactics to be by example. We take different types of knowledge that is produced from literature. For example, we take the examples of creating methods from Dittrich et al. \cite{dittrich2008cooperative} while we take examples of creating products from Staron et al. \cite{ochodek2022chapter}. 

    The most important for the students is to understand that we can create different types of knowledge in action research. We must explain that action research is flexible, but that knowledge production must be planned in advance, otherwise it is not systematic and can be prone to problems with quality, replicability or reliability.
\end{teachingenv}

\section{Phases of Action Research}
\label{sec:phases}
\begin{figure}[!htb]
    \centering
    \includegraphics[width=\textwidth]{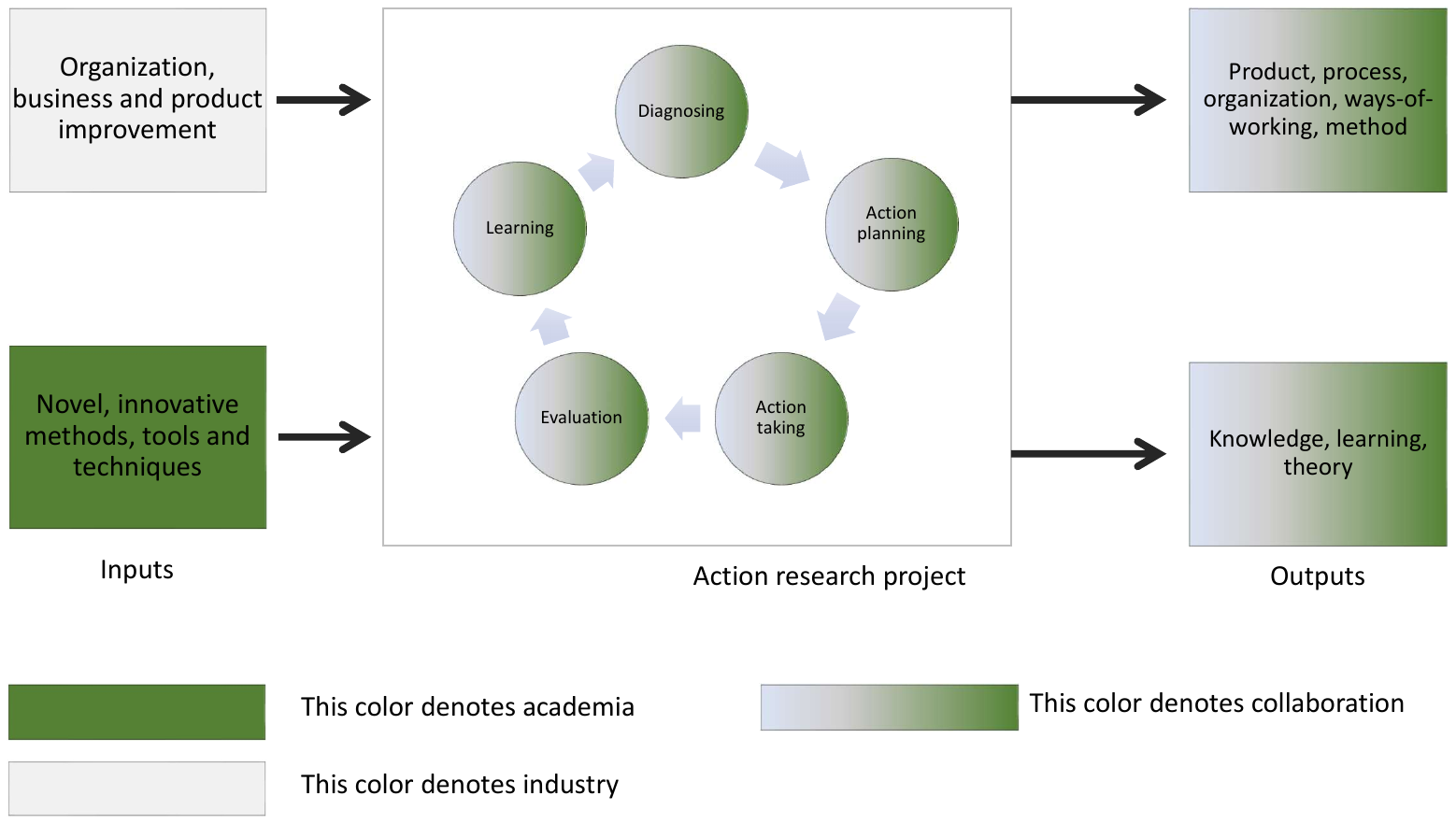}
    \caption{Phases of action research}
    \label{fig:action_research}
\end{figure}

\noindent
Action research is a cyclic methodology with several phases in each cycle. The number of phases can differ, depending on different schools of action research. In software engineering, the most common is the five phases cycle, depicted in Figure \ref{fig:action_research}

\textbf{In the diagnosing} phase we start with addressing the question of \emph{What is the real
problem that we should address in this cycle?} Although the question is often partially answered when initiating the project (or from the previous cycle), it’s important to specify which part of the problem should be addressed in each cycle.

The first phase of each action research cycle --diagnosing -- is unique for action research. Instead of starting a project with a detailed problem formulation, action research recognizes the fact that one needs to be embedded in the context in order to elicit the problem correctly. Therefore, every action research cycle starts with a precise diagnosis of which problem should be solved. Action researchers should start by collecting opinions and symptoms which they need to explore in order to decide which challenge to address during the action research cycle. It’s important that the researchers focus on discussions with the practitioners when exploring the context and deciding what to do. The problem to be solved in each cycle should be limited in scope, and its effects should be measurable (or at least observable).

Melin and Axelsson \cite{melin2007action} recognize two types of identifying research problems:
\begin{enumerate}
    \item when an action researcher identifies the problems, i.e., research-driven initiation, or
    \item the problems are presented to the action researcher, i.e., problem-driven initiation.
\end{enumerate}

From my experience, the first type, i.e., research-driven initiation is more common for the diagnosing part, whereas the problem-driven initiation is more common for the overall definition of the research project. Avison et al. \cite{avison2001controlling} recognize the possibility of both parties working together in recognizing the research problem, i.e., collaborative research initiation. 

\smallskip
\begin{subtitleenv}
In my work, one of the first action research projects was initiated by the company -- problem-driven initiation \cite{staron2012release}. The company wanted to improve predictability of the release dates of their product. I was involved in another project, where my role was to design a measurement program. 

Since the problem was relevant for my project and it was relevant for the company, we started working on it. We called it the \emph{Release Readiness project}. The diagnosing phase led us, later, to a more detailed definition of the problem and solving it. 
\end{subtitleenv}
\smallskip

\textbf{In the action planning} phase, actions/interventions are planned in a single cycle is always done in a collaborative manner. Academic researchers, industrial researchers, and practitioners need to work together to decide who does the actions and when.

The collaborative nature of the action planning phase provides a unique opportunity for both practitioners and researchers to engage in discussions. The discussion are often aimed at finding ways to solve the problem diagnosed in the first phase and identify resources, products, and processes to be investigated and adjusted.

In the action planning activity, the action team (which is how I call the research team) discusses their plans with the reference groups and needs to get approval for the required resources from the management team. The plans need to be aligned with theoretical foundations of the work, i.e., the action team needs to identify theoretical or empirical work relevant for the diagnosed problem and plan the actions accordingly.

In this phase, the action team, together with the reference team, makes the plans for which data should be collected, from which objects, using which tools. The team also plans which analysis methods should be employed to assess whether their actions lead to solving the diagnosed problem.

Often, although far from always, the action team plans their actions using standard project planning tools, like Gantt charts and work breakdown structures. However, these are often lightweight and documented only internally for the action team to follow and use as a communication tool to management.

\smallskip
\begin{subtitleenv}
In my release readiness project, action planning was done in a form of a workshop. Together with the stakeholder (manager responsible for the release), the action team and the reference team, we brainstormed which metrics should be used for our predictions. During this workshop, we identified and discussed various relationships between the metrics in order to capture the relevant empirical phenomena that impact the release -- e.g., test progress, current quality of the product. 
\end{subtitleenv}
\smallskip

\textbf{In the action taking} phase the team conducts the interventions -- they change the processes, tools or methods of the host company in order to systematically collect data. 

The phase is executed according to the plans laid out in the previous phase and is conducted by the action team. The reference group is involved on a regular basis to provide feedback and to help the action team to solve the challenges that they encounter \cite{antinyan2016validating}.

The action taking phase is specific for action research as it is one of the research methodologies where making changes are allowed, midst in the operations. It’s called a flexible research design methodology \cite{robson2016real}. For example, the action team is allowed to change the ways of working for software development teams and observe these changes. It is important to note that the action taking phase is both about making the change and observing its effect. As action research is a quantitative methodology, the data collection activities provide the possibility to reduce the bias of subjective observations and provide quantitative evidence. This quantitative evidence is used in the next phase -- action evaluation --to assess the results of the action/intervention.

\smallskip
\begin{subtitleenv}
This phase of our release readiness project gravitated around collection of data, calculation of metrics/indicators, presentation to the management and adjusting resources in the project. We collected data weekly and presented them to the stakeholder. The stakeholder made the decisions about the potential re-allocation of resources to fulfill the organization's goals. 
\end{subtitleenv}
\smallskip

\textbf{In the evaluation} phase the action team analyzes the data collected from the previous phase. The team uses statistical methods to make the analyses and presents the results to the reference team and the management.

In case when the data shows that the diagnosed problem is indeed solved using the actions taken, the outcome is straightforward. If the data is inconclusive, the action team either needs to plan for additional analyses and additional data to be collected or needs to pivot, i.e., finalize the current cycle, specify learning, and find a new diagnosis of the problem given the new data collected. Then the action team continues with the next cycle to address this diagnosed problem.

In the evaluation phase, the action team could either take the quantitative approach -- use the same statistical methods as experimentation, i.e., descriptive and inferential statistics --- or the qualitative approach -- use grounded theory, thematic analysis or workshops. The action team also needs to assure that the analysis of their data is aligned with the theories used in the cycle. This is important in order to make the contribution to the theory-building in the next phase.

\smallskip
\begin{subtitleenv}
To evaluate the release readiness indicator, which is how we called the metric developed in the project, we collected statistics of decisions made during the project. We kept track of all changes that the stakeholder took and how they impacted the organization, the product and also the release readiness indicator (and the metrics used to calculate it). We conducted a workshop in the same group as for the action planning and complemented the quantitative metrics with the qualitative assessment of the results of the project. 
\end{subtitleenv}
\smallskip

\textbf{In the specifying learning} phase the documentation and generalization of learning takes place. It is done both as practical guidelines for the involved organizations and contexts and as theory-building for the research community.

The practical guidelines are often specified in terms of guidebooks, white papers, and instructions at the company’s web. For example, software development teams often use wikis to specify good practices and document good examples. That’s often when the results of action research cycles can be found.

The contribution to the theory-building is often specified as scientific papers, with the scientific rigor and relevance. It is often the case that these are documented as experience reports from industrial studies \cite{ochodek2022chapter}.

\smallskip
\begin{subtitleenv}
In the case of our release readiness project, we presented the results to a larger audience on an annual conference of the entire company. We also published a paper and presented it on an international conference \cite{staron2012release}. We prepared internal information for other stakeholders who adopted this method for other parts of the organization and other products.     
\end{subtitleenv}
\smallskip

\section{The research team == the action team}
\label{sec:teams}
\noindent
Action research can be conducted by researchers within the host organization or the natural environment of the participants. In fact, this is how the concept of action research appeared \cite{lewin1946action}. However, this is not how action research is conducted in software engineering. In software engineering, we usually combine academic researchers with industry professionals. In this context, we can refer to this group as the \emph{action team} -- which is the research team that includes (academic) researchers and (company) practitioners. I deliberately use brackets for academic and company, as the boundaries can be fuzzy. For example, an academic researcher may be employed by the company for the duration of the project in order to handle the IPR (Intellectual Property Rights) in a simple way. 

\begin{figure}[!htb]
    \centering
    \includegraphics[width=\textwidth]{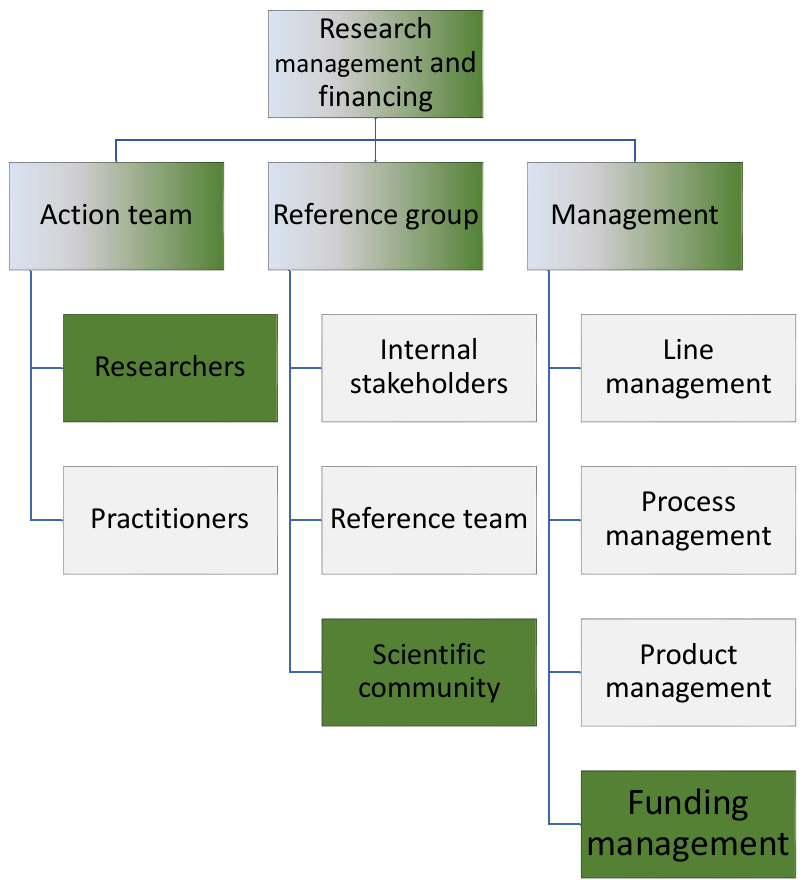}
    \caption{Action team in the context of other teams. The action team is the one conducting the research. The reference team is the team who helps in evaluating and scaling-up the work of the action team. The management team holds the responsibility for the financing, staffing and monitoring the action research study.}
    \label{fig:action_team}
\end{figure}

Here, we focus on the typical software engineering context -- where the researchers and practitioners collaborate in a joint project. In this case, the action team consists of two roles and is surrounded by other stakeholders -- the reference team and the management team, as presented in Figure \ref{fig:action_team}. 

In the \textbf{action team}, researchers and practitioners have different roles in the action team, despite the fact that they work in the same team. 

The \textbf{practitioners' primary task} is to work with the interventions. Since the practitioners are working at the host organization, they can conduct the interventions/actions. They can change their ways of working and provide data through these changes. They can (and should) participate in the action planning, evaluation and learning too, but their expertise is crucial for the interventions. The practitioners bring in the following to the action team:
\begin{itemize}
    \item Internal competence about the company's processes, ways of working and decision structures. 
    \item Possibility to use tools and information from inside of the company, which can require employee-only access.
    \item Historical knowledge about the company's processes, products and strategies. 
    \item Possibility to internally champion the action research study, and in particular the action/intervention. 
    \item Possibility to invite others to join the relevant parts of the action research cycle, e.g., bring in additional software developers during the evaluation of the outcome of the intervention. 
\end{itemize}

The \textbf{primary task for the researchers} is to prepare plans for the actions/interventions, observe how they are conducted, collect the data and analyze it. They can (and should) participate in the interventions, but since they are often not part of the host organization (at least not to the same extent as the practitioners), they cannot conduct the interventions alone. The researchers bring in the following to the action team:
\begin{itemize}
    \item External (to the host organization) competence about the newest research results, tools and technologies. 
    \item Possibility to test tools outside of the organization before the action team decide (during action planning). 
    \item Systematic planning of data collection -- during baseline (action planning, before action taking), during the intervention (during action taking) and after the action taking. 
    \item Objective methods for analysis of the data collected during the action research cycle. 
    \item Possibility to replicate the study at other organizations (or even open source products/teams) to study the generalizability of the results obtained at the host organization. 
\end{itemize}

The role of the \textbf{reference group} is to provide the possibility to get feedback on the progress of the project and to reduce biases. The reference group also helps the action team to diagnose the problems and therefore steers the project in the right direction. As the action team is conducting the research, they are biased toward a positive outcome of the project. The reference team is responsible to provide the action team with the feedback on how to reduce this bias and identify when the bias is jeopardizing the outcome of the study.

Finally, the \textbf{management} of the company is important as they decide upon the resources needed for the project. The resources, in turn, determine the scope of the project. The product and process management are important as they help to support the project in making the right impact of the results of their actions.

\begin{teachingenv}
    Teaching team dynamics is beyond the scope of this chapter, but it is crucial for long-term collaboration.

    When teaching collaboration, it is important to consider both academic and industrial perspectives. From the academic perspective, we must discuss the quality of research, including publication venues, targets, and academic goals. This involves understanding how to produce high-quality research that meets academic standards and contributes to the body of knowledge in the field.

    From the industrial perspective, it is essential to discuss the practical implications of implementing technology. This includes understanding how to scale a studied technology from a small-scale application within the host team to broader use in other teams and organizations. It is important to differentiate between research prototypes and fully-developed software engineering tools and methods. This discussion should cover the challenges and strategies for transitioning from experimental research environments to real-world applications, ensuring that technologies are robust, scalable, and ready for industry adoption.
\end{teachingenv}

\section{Host organization}
\label{sec:host}
\noindent
Selecting the right host organization has a crucial effect on the success or failure of action research studies. Although there is no universal template for a suitable organization, there are several characteristics that are important for successful action research collaboration. 

The first and foremost condition for the success of the action research project is \textbf{the organization's willingness and ability to implement interventions}. Even the best research result is not very useful if it is not implemented at a company, which is the core of action research studies. In the context where the collaboration happens across borders of two entities -- university and company -- we can achieve this in a few ways:
\begin{itemize}
    \item Researchers spend a significant time at the company. They are part of the host company operations regularly spending from two to five days a week at the company. This creates the possibility to create a real team spirit and create an atmosphere where questions can be asked (and answered) directly and immediately. 
    \item The practitioners who are part of the action team are communicating daily with the researchers (and vice versa). In the post-pandemic times, online collaboration tools can be a great ways of quick communication. Daily, quick, and agile check-in meetings help to both move the project forward and to create the atmosphere of transparency, trust and being one team. 
    \item The action team has weekly meetings where the project is discussed and followed up. The stand-up meetings are a bit orthogonal to the previous two, but they are the least that should be done in the project to ensure that things are moving forward. 
\end{itemize}

Naturally, the best option is the first one, where the researcher can participate in the operations of the company. Let me provide an example why this was important in my career. 

\smallskip
\begin{subtitleenv}
When I do action research studies, I aim for the first option -- being part of the host organization's operation. Let me bring up one experience where I worked with one company on developing measurement systems \cite{staron2009using, staron2018industrial}. In this collaboration, the host company asked me to spend 50\% of my time on their premises. Although it sounded strange at the beginning (and foreign to my academic colleagues), this turned out to be a great way of co-creation of research and innovation.

During the time of my stay, I could immediately ask my industrial colleagues about the practices which we focused on in our research, and they could learn what kind of research questions and methods are important. One example of my question was \textit{What's the difference about a defect found in a system test vs. a defect found in a unit test?}. An example of my colleague's question was \textit{Why do we need to formulate our research question "To which degree can we predict defects?" and not "Can we predict defects?"}

Although the above questions seem trivial, we needed to ask them to learn. Only sitting together for a significant part of our work enabled (and encouraged!) us to ask these kind of questions to one another. In the long run, asking questions leads to increased trust and to more fruitful collaboration. 
\end{subtitleenv}
\smallskip

\textbf{The organization provides the research team with access to the information} which is needed to conduct the action research study. The action team must be able to collect data about their interventions from source systems and therefore the access to the information should be unrestricted. In practice, this means that everyone in the action team should be able to have the same access, for the information relevant to the project. There are a few ways to enable that:

\begin{itemize}
    \item The researcher uses company's infrastructure for the project in terms of the computer equipment and IT accounts. The researcher must sign the relevant Non-Disclosure Agreements and other legal documents. The same is true for her/his university. The information that is stored at the company's infrastructure is legally under the company's jurisdiction and therefore the researcher does not have to consider additional security. Analyses, notes and pre-published materials can be safely stored at the company until they are assessed to be made public. This option allows the action team to work unrestricted until they need to prepare a publication, then the relevant management at the company can provide the team with the feedback on what can (and cannot!) be published. It is also easier for the researcher to understand what is possible to published based on the daily work, which often leads to a smooth approval process at the company side. 
    \item The host company agrees up-front which information is relevant for the project and shares that information. If information cannot be publicly shared, then the practitioners (who are part of the action team) must actively analyze it, discuss in the team and then get it approved. This can be achieved when different participants of the action team have different roles in the team -- the practitioners must take more responsibility for the research part. In practice, this means that the analyses are done on the computers of the company, but the researchers can provide an open version of the toolkit, for example by receiving mock-up data in the right format. 
\end{itemize}

The ways of accessing the information is not as important as the information and collaboration itself. 
\smallskip
\begin{subtitleenv}
In my work I usually try to ask for the company's infrastructure for my action research projects. When being part of the company's team (as I mentioned in the previous box), this is both easier for the company and for me. I'm ensured to be able to use the information when I prepare for actions (diagnosing and action planning) as well as after the action (evaluation and learning). 

My colleagues do not have to constantly consider the information's security class and when we prepare our publications, we can keep the managers informed and "in-the-loop" so that we know which information can be sensitive. I also prepare a so-called open replication package where we create a mock-up data (or collect it from a relevant open source repository at GitHub) in order for others to use our research. 
\end{subtitleenv}
\smallskip

The organization has \textbf{trust in the action team and is therefore transparent} in the collaboration. Trusting someone is naturally harder than it sounds. It is a combination of both the skills and the so-called personal chemistry. Not everybody can be a good action researcher and not everybody can be a good action research practitioner. However, there are a few good practices that help to develop trust in the team. 

\begin{itemize}
    \item The action team is the researcher's priority. They should consult the team before seeking academic advice. They should also communicate frequently and openly with the team, even about minor issues. Asking questions shows curiosity and builds trust, even if the question seems simple and silly.
    \item The team should honor their commitments. They should follow through with the agreed activities and report on their outcomes or deviations from the planned outcomes. Sometimes trying is worth as much as succeeding, as we are in the context of research and knowledge production. Sometimes, learning that something does not work is even more important than forcing a success.
    \item The team should respect the reality of the host organization. They should acknowledge the diversity and complexity of the people, products and processes involved in the action research project.
    \item The team should learn from their mistakes. They should not hide or ignore errors, but rather discuss them openly and constructively. Transparency and accountability are key to effective work in the action team. 
    \item The team should treat everyone with respect. They should recognize that working with companies means working with human beings, not abstract entities. They should appreciate the goals, schedules and abilities of the people in the organization, and not impose unrealistic expectations or demands.
\end{itemize}


\smallskip
\begin{subtitleenv}
Learning from mistakes is crucial. How we admit and fix them matters most. As a senior researcher, I still make errors sometimes. For instance, I once left an index column in the analyses, which gave great results. But when I checked them, I realized my mistake and told my colleagues. It was hard because they trusted my skills and the results. But when I confessed and showed the worse results, we became more trusting, published the paper and still collaborate.
\end{subtitleenv}
\smallskip

We must know and respect the people in the organizations, as they have their own goals, schedules and abilities. Part-taking in an action research study means interest and willingness to collaborate, not a legal contract to deliver. 

\begin{teachingenv}
    When teaching the aspects related to the host organization, we need to focus on selecting the host organization. In particular, discussing the criteria for when a collaboration can be profitable for both the academia and industry. In order to do that, we can study guidelines for industry-research collaborations \cite{staron2011factors, wohlin2021guiding}.

    We also need to teach how to conduct daily communications and meetings. For example, we can simulate this by conducting these meetings during every course meeting. We can ask each of the participants to report on their activities, related to the course, during the past week.  

    We must discuss the integration of the research activities into the operations. In particular, we should explain that action research must contribute to the industrial practice; we should also discuss methods for identifying problems to study at the host organization. 
\end{teachingenv}

\section{Interventions}
\label{sec:interventions}
\noindent
Although the research team and the host organization are extremely important, it is the intervention (also called action) that is in the core of action research projects. The intervention is important as it is what gives origin to the knowledge that we generate through evaluation and learning later on. We need the interventions in order to understand how the organization reacts to the changes that are introduced in the project, as well as to collect data to analyze. This could indicate that an intervention could be any kind of activity involving the host organization, but it's not that simple. 

A good \emph{intervention is when we introduce a change in the host organization in a systematic and planned manner which has impact on its operations}. The key elements of this definition are:
\begin{itemize}
    \item \textbf{Planned} intervention, which means that the action team must prepare the action and the host organization must be prepared for that action. We must also choose the right timing for the intervention. 
    \item \textbf{Systematic} intervention, which means that we need to establish infrastructure for collecting data from the intervention. The action team must collect data before, during and after the intervention. Before the intervention, the team must establish a baseline which is used for comparison with the data collected after the intervention. During the intervention, the action team must collect the data about the impact of the intervention on the operations. 
    \item \textbf{Impactful} intervention, which means that it has observable effects on the host organization. The impact can be in terms of change or the ways of working, introduction of a new tool or using new type of data in decision-making processes (for example). The key is that it is part of the processes in the organization, not alongside. 
\end{itemize}

Although it becomes clear what a good intervention is over time, we can provide a few good and bad examples of interventions. 

Good examples:
\begin{enumerate}
    \item Using a new tool to the company operations. When we introduce the tool, we make change in the host organization's processes. We can establish a baseline before the introduction of the tool, observe the introduction itself and observe the effects of the introduced tool afterwards. 
    \item Using new information in a process of making a decision. When we use the new information, we change the process of making the decision. We can establish a baseline before, observe how the new data is used in the process and observe the effects of the decision afterwards. 
    \item Changing the process of handling tickets -- migration from e-mails to JIRA tickets. As it is a clear change, we can establish a baseline before, observe the change and then observe the effects afterwards. 
\end{enumerate}
\bigskip

\begin{subtitleenv}
One example of the intervention from one of my studies is using analysis of the source code to improve coding guidelines \cite{ochodek2022chapter}. The analysis of the source code was the preparation for action taking and the actual action taking was the use of the results in the organization. The effect was the improvement of the coding guidelines at the company. 
\end{subtitleenv}
\bigskip

Bad examples:
\begin{enumerate}
    \item Presentation (but not use) of results for the host organization. It is a bad example because it does not guarantee that the presentation has impact (presentation is not the same as using the results).
    \item Development of a new tool for the host organization. As long as the tool is not used, it is not an intervention. 
    \item Analysis of data at the host organization. Just like developing a tool, analyzing the data is not an intervention until the analysis is used at the host organization. 
\end{enumerate}

The two lists above indicate the trend that is important to take note of -- a good intervention introduces a change. The action team must prepare for such an intervention by establishing a baseline. 

The baseline for each intervention has to be related to the intervention itself. The action team must plan for collecting data that can be compared to after the intervention. Since the intervention cannot be undone, the action team can collect the data for the baseline only \textbf{before} the intervention. 

The data collected during the intervention should also be related to the intervention, and it should be complemented with interviews and observations. The latter are important for capturing such effects that the action team did not plan for. For every intervention there can be unpredicted consequences and these need to be captured; the best methods for capturing them are interviews and observations. 

Finally, after the intervention the action team must collect data that related to the baseline, consequences identified during the intervention and finally the effects of the intervention. The latter must be done in order to prepare for evaluation and learning from the intervention itself. Oftentimes, the lessons' learned from the intervention go beyond the initial expectations -- that's the power of action research!
\bigskip

\begin{subtitleenv}
In one of the projects, we introduced a new defect prediction model as a means to prioritize testing and development resources \cite{staron2010defect, staron2010method}. In order to evaluate the model, we collected data about resource allocation before the model was used. We interviewed program managers and integration specialists about their work with resource allocation and defect management. We also collected data about the weekly (and monthly) trends of defect inflow. 

During the introduction, we met with the stakeholders (program manager and integration specialist) once a week to follow up on their activities and the impact of the action. We showed them the defect predictions and asked whether these predictions deviated from theirs. We also asked about which actions they take to ensure that the defect inflow does not get out of control.

Once the introduction was complete, we collected new data in the same way as the baseline data -- trends of defect inflow and interviews. 
\end{subtitleenv}
\bigskip

I often use the following checklist when designing the intervention (yes/no answers only):
\begin{enumerate}
    \item Is the plan of the organization to adopt the intervention as part of their operations if it is successful?
    \item Does the intervention/action make a change in the host organization?
    \item Can we collect data (quantitative or qualitative) about the baseline situation before the intervention?
    \item Can we be part of the intervention to collect the data?
    \item Do we have mandate to adjust the intervention in case of unpredicted events?
    \item Can we collect the data after the intervention to observe the effects of the intervention?
\end{enumerate}

If I can answer at least four out of these six questions positively, then I know that I can proceed with the intervention. The other two usually fall into place once I start. 

Action research is a cyclic methodology that allows us to repeat series of phases several times. We diagnose the problem, plan for the intervention/action to address the problem, conduct the intervention, collect and analyze the data and learn from it. Therefore, we often have the possibility to make relatively small interventions in each cycle, but accumulate changes over several cycles. In one cycle we learn from the intervention and plan for new ones. The small interventions in each cycle have a number of advantages, in particular they do not require radical changes in the host organization. They also allow to adjust or pivot between one cycle and another if we discover and learn new facts. 

\begin{teachingenv}
    Since interventions are the most distinct characteristics of action research, we need to spend a significant effort to teach them. 

    I recommend to have a workshop-based approach to teaching interventions. First, start with the definition of the concept of the intervention, then focus on aspects related to the impact of the intervention and then focus on how to behave in such a project -- explore existing codes of conduct at the companies, conflict management and expectations management. 

    This can be achieved by a set of exercises:
    \begin{itemize}
        \item Discuss the definition of an intervention/action -- focus on discussing the contrast between making an intervention and studying a case (case study); discuss the difference to participatory observations from the perspective of the impact and bias in the organizations. 
        \item Take up one of the examples of good interventions and one counter-example; discuss their impact on the organization and the scientific community.
        \item Take up a case of a conflict at a workplace (from literature) and discuss how to handle conflicts; it is particularly important for younger researchers who need this kind of training for their future careers. 
    \end{itemize}
\end{teachingenv}

\section{Ethics of Action Research}
\label{sec:ethics}
\noindent
The main principle of ethical research should always be to \emph{do no harm}. Our actions as researchers should always have the goal to improve our society and not to harm it. In action research projects, we must protect both the individuals who take part in the research and their organizations. The host organizations invite researchers to become part of them and they trust them that they will bring no harm to their employees or management. Therefore, it is very important that open climate, questions and discussions are the best way to address ethical issues in the action research projects. 

Since action research projects are based on interactions with its context and software engineers, we need to provide ethical considerations for the project, in particular, how we will select the participants, how we will store their personal data, and how we anonymize the data so that it does not lead to any harm to individuals and organization.

An interesting aspect is the legal part of the collaborations with the companies. We need to make sure that we have all the agreements in place and that all intellectual property rights are handled according to the regulations specific to the countries where the research is conducted. Avison et al. \cite{avison2001controlling} refer to this as a formalization of the research project.

\textbf{Selection of participants in our studies} must be done using ethical principles. First, we must recognize and understand bias in this process. Although we may be given initial team to work with, we should strive to increase diversity in the project overall. For example, if we have male-only action team, we should find female colleagues in the reference team to capture a diverse set of view, opinions and experiences in our research project. 

\textbf{Handling conflicts of interests} must also be recognized and monitored during the entire progress of the action research project. Despite the best efforts to be unbiased, there is always a risk of the researcher bias and the Hawthorne's effect. The researcher bias is the inability to objectively scrutinize research results. The action team's stake in the project is often to improve and therefore the action team is inherently positively biased and must use reference teams and the management team to counter-balance it \cite{staron2020action}.

\textbf{Handling confidentiality} can be seen as conflicting the principles of academic freedom and the science's ability to serve greater good. However, in practice, this is not the case. There are two levels of confidentiality -- within the company and externally to it. Internally, the team should be open as they are part of contractual agreements. The external confidentiality refers to what we are obliged to keep confidential about the company, its processes, products and intellectual property rights. We can always generalize our observations, evaluate their validity in other contexts (e.g., open source communities) and discuss on a general level. We must remember that the scientific community is not interested in so-called "whistle-blowers" from the companies, but in new knowledge. This is also what all host organizations are interested in. Whenever we are in doubt, we can always consult company's management, legal team, university's legal team or even university's ethical boards regarding confidentiality. 

\textbf{Overpromising} is also one of the aspects which is important to take into consideration when conducting action research projects \cite{morton1999ethics}. It is easy for the researchers to overpromise to deliver in an action research project and then continue to try to push too much to deliver on that promise. However, the action team should define their deliverables upfront and should be prepare to adjust the plans over time. 

\textbf{Quality of research} is sometimes seen as being in conflict to delivering to the company \cite{morton1999ethics}. The action team may feel pressure of delivering new value to the host organization on the cost of quality of research. The quality of research is very cost-intensive as it requires validations, checking for confounding factors and replications, which are often not as value-driving as diagnosing, addressing and validating new problems. Involving other organizations is often much more demanding and time consuming, but increases the quality of the research. 

In addition to these ethical concerns, there are others, which the action team needs to discover themselves. The fact that they are embedded in the host organization means that they are inherently biased and therefore then need the reference team and the management team to discuss and vet ideas and results. I strongly encourage the action teams to discuss potential conflicts of interest based on their specific context. 

\begin{teachingenv}
    The best way to teach the ethics of action research is to start from studying ethical guidelines that apply. Every country, company and university operate in a different legal space and therefore the action team must:
    \begin{itemize}
        \item Study and understand the contract between all parties in the collaboration. 
        \item Identify and operationalize the ethical guidelines applicable, e.g., for using human subjects in case studies. 
        \item Engage in discussion about the legal aspects of ownership of research results -- include company and university legal advisors. 
    \end{itemize}
    \noindent
    Just by reading, exploring and asking critical questions about the meaning of clauses in legal documents, NDAs (Non-Disclosure Agreements), and ethical review boards guidelines, we raise awareness of these issues. The students can be asked to identify and explain some of these principles for their project or a project that they find in a literature. 
\end{teachingenv}

\section{Guidelines for teaching action research}
\label{sec:guidelines}
\noindent
Engaging in action research is a very rewarding way of conducting research studies. It allows us to be more embedded in the reality of software development organizations, products, processes and markets. Getting to the point when the action team understands the intricacies of action research requires training. Teaching action research, therefore, should revolve around learning the following elements. 

\begin{enumerate}
    \item Epistemology of knowledge in action research. First, and the foremost, new action researchers needs to understand what kind of knowledge is produced in action research and therefore what to expect from it. 
    \item Interventions. Once we introduce the type of knowledge that is created in action research, the researchers need to learn what a good and bad intervention is. They need to understand how the knowledge is produced and therefore what we are expected to do in the action research projects. 
    \item Action team. The interventions need to be conducted by the team and therefore the team needs to learn about the roles and responsibilities of each member of the action research. The action team must learn about the differences in roles and the expectations. This makes the team understand how the knowledge is created in action research. 
    \item Host organization. Once the team understands what kind of knowledge is created and how, in action research, the team needs to understand their obligations towards the host organization and how to navigate the challenging balance between operations and research. 
    \item Planning and executing action research cycles. Now, the action team is ready to learn about the phases of the action research. Each phase needs to be explained and discussed with the action team. 
    \item Deliverables. Providing new knowledge has to be done in some form. Therefore, the action team needs to be educated in what kind of deliverables are expected and how to assess if a given deliverable has a good quality or not. 
    \item Collaboration. Since the action research is collaborative in nature (at least the one that we describe in this chapter), the action team must understand how to collaborate. Models and methods for assessing team maturity are very useful for it, e.g., \cite{meding2021meteam}. 
    \item Ethical aspects of action research. In order to understand what the action team can, could, should and should not do, the action team must learn about the ethics of action research. They need to understand how to include diverse perspectives in research, how to cause no harm and how to balance academic freedom with the obligations towards the host company and the research project. 
    \item Contrasts to similar research methodologies. Finally, the education in action research should include learning about the contrasts to other, related, research methodologies. The action team must, in particular, understand the differences between the action research and case studies, experiments and design science research.  
\end{enumerate}

Naturally, action research studies can be conducted with different focus in software engineering. Some studies can be more focused on the technology development, e.g., understanding how to use generative AI to create tests, while some studies can be focused on social aspects of software engineering, e.g., how to create a better performing team. Therefore, it is important to recognize the focus, the context, of action research studies and to complement the education in action research with the education in that context.

\section{Summary and conclusions}
\label{sec:conclusions}
\noindent
Action research is an applied empirical research methodology which focuses on direct interventions, actions, conducted at industrial partners. The methodology has been been successfully used in medicine, nursing, education and social sciences. However, its adoption in software engineering is just starting to take off. 

In this chapter, we learned what action research is and how it is conducted. We have also learned about the pillars of it -- the action team, the host organization and the interventions. We also learned about the ethics of action research and the way in which we can provide value to both the academic community and the industry. Finally, we finished up this chapter with guidelines on how to teach action research. 

As further reading, I recommend to read the newest research studies using action research in software engineering, as this is the kind of research methodology that is on the rise.

\begin{acknowledgement}
This book chapter is partially sponsored by Software Center (\url{www.software-center.se}) which is a collaboration between five universities and 17 companies. I would also like to thank Dr. Birgit Penzenstadler and Prof. Yvonne Dittrich for very useful comments on the drafts of the chapter.
\end{acknowledgement}

\bibliographystyle{acm}
\bibliography{software}

\end{document}